\documentclass[aps,pra,reprint,amsmath,amssymb,superscriptaddress,longbibliography]{revtex4-2}

\usepackage[utf8]{inputenc}
\usepackage[T1]{fontenc}
\usepackage{graphicx}
\usepackage{bm}
\usepackage{hyperref}
\usepackage{xcolor}
\usepackage{mathtools}
\usepackage{physics}
\usepackage{siunitx}
\usepackage{booktabs}

\newcommand{\FI}{F_{\mathrm{QD}}}
\newcommand{\HS}{H_S}
\newcommand{\Hol}{\chi}
\newcommand{\calH}{\mathcal{H}}

\newcommand{\kB}{k_{\mathrm{B}}}

\begin{document}

% ===== Title =====
\title{Functional Information in Quantum Darwinism:\\
An Operational Measure of Classical Objectivity}

% ===== Authors =====
\author{Arda Batın Tank}
\email{ardabatin.tank@std.yeditepe.edu.tr}
\affiliation{Department of Physics, Faculty of Engineering and Natural Sciences, Yeditepe University, 34755 Istanbul, Turkey}

% ===== Date =====
\date{\today}

% ===== Abstract =====
\begin{abstract}
Quantum Darwinism explains the emergence of classical objectivity through the redundant encoding of pointer information in environmental fragments. However, existing diagnostics rely on arbitrary thresholds or structural assumptions that limit their operational applicability. We develop a framework based on \emph{functional information}, $\FI(\delta) = \log_2 R_\delta$, which quantifies objectivity as the abundance of environment fragments that individually carry at least $(1-\delta)H_S$ bits of classically accessible pointer information, as bounded by the Holevo quantity. Using onset statistics rather than parametric fits, we extract redundancy $R_\delta$ from the fragment size at which adequacy becomes typical. Simulations of a heterogeneous pure-dephasing model reveal three robust features: rapid early-time growth of $\ln R_\delta$, smooth crossover to saturation, and capacity-limited plateaus at $\FI^{\mathrm{plateau}} \lesssim \log_2 N$. We establish thermodynamic constraints showing that each additional bit of $\FI$ doubles the minimal heat dissipation required for record stabilization. These results frame classical objectivity as a quantifiable, resource-limited phenomenon.
\end{abstract}

% ===== Keywords =====
\keywords{quantum Darwinism, decoherence, objectivity, functional information, Holevo bound, quantum-to-classical transition}

\maketitle

% ===== Introduction =====
\section{Introduction}
\label{sec:intro}

The emergence of a stable, intersubjectively agreed-upon classical world from quantum mechanics remains a central foundational challenge. While quantum theory permits arbitrary superpositions, everyday experience presents definite outcomes upon which independent observers reliably concur. Decoherence theory addresses part of this puzzle by showing how environmental monitoring selects a preferred pointer basis through einselection~\cite{Zurek2003,Schlosshauer2007,Schlosshauer2019}. Yet decoherence alone does not explain \emph{intersubjective agreement}: why do spatially separated observers, accessing disjoint portions of the environment, independently arrive at the same classical description?

Quantum Darwinism (QD) addresses this question by reconceptualizing the environment not merely as a source of noise but as a \emph{communication channel} that broadcasts information about the system's pointer observable into multiple fragments~\cite{Zurek2009,BlumeKohoutZurek2005,OllivierPoulinZurek2004}. In this framework, objectivity emerges when many observers can independently extract nearly complete pointer information from their respective environmental fragments without disturbing the system or coordinating with one another. The key signature is \emph{redundancy}: the same classical information is encoded many times over in the environment. Recent work has demonstrated that such redundant encoding can emerge generically under broad conditions~\cite{Brandao2015,Knott2018}.

Several diagnostics have been developed to detect and quantify this redundancy. Partial information plots track how the quantum mutual information $I(S\!:\!F)$ grows with fragment size, identifying a ``plateau'' where adding more environment yields diminishing returns~\cite{BlumeKohoutZurek2006,Riedel2010,Zwolak2014}. However, extracting a single redundancy scale from such plots typically requires imposing \emph{ad hoc} thresholds---e.g., declaring a fragment adequate when it captures 95\% of $H_S$---whose physical justification remains unclear. Spectrum broadcast structure (SBS) provides a mathematically precise, threshold-free criterion: the joint system-environment state must factorize into a classical-quantum form with orthogonal conditional states~\cite{Horodecki2015,Korbicz2021}. While elegant, SBS is often too stringent to verify beyond small systems and may exclude operationally objective scenarios that fall short of perfect orthogonality~\cite{Roszak2019}. Quantum mutual information captures total correlations but conflates genuinely redundant classical records with residual entanglement, potentially overstating objectivity~\cite{Le2018,Le2019}. Recent studies have also explored QD beyond simple spin-environment models, revealing richer phenomenology~\cite{Zwolak2017,Ciampini2018}.

These limitations motivate a \emph{conservative, operational} approach. Rather than asking how much total correlation exists, we ask: \emph{How many environment fragments are, each on their own, good enough to serve as reliable classical records of the pointer value?} This question shifts the focus from correlation measures to \emph{adequacy counts}---the abundance of fragments that clear an operationally meaningful bar.

In this work, we develop a framework based on \emph{functional information in Quantum Darwinism}, denoted $\FI$. The concept adapts the notion of functional information from origins-of-life studies, where it quantifies the rarity of molecular configurations capable of performing a specified function~\cite{Hazen2007}. In QD, the ``function'' is information storage: a fragment is \emph{adequate} if it permits an observer to infer the pointer outcome with high confidence. We operationalize adequacy through the Holevo bound~\cite{Holevo1973}, the fundamental ceiling on classically accessible information from a quantum ensemble. A fragment $F$ is deemed $\delta$-adequate when its Holevo quantity satisfies
\begin{equation}
\Hol(\Pi_S\!:\!F) \;\geq\; (1-\delta)\,\HS\,,
\label{eq:adequacy}
\end{equation}
where $\Pi_S$ is the pointer POVM, $\HS = H(X)$ is the Shannon entropy of pointer outcomes, and $\delta \in (0,1)$ is a tolerance parameter. The one-sided, capacity-aware nature of this predicate ensures conservatism: only fragments with genuinely accessible classical information qualify.

Rather than fitting parametric curves to information-theoretic quantities, we extract redundancy from \emph{onset statistics}. For a given fragment size $m$, we compute the fraction $\Phi_\delta(m)$ of randomly sampled fragments that satisfy Eq.~\eqref{eq:adequacy}. The \emph{onset} size $m_\delta^\star$ is the minimal $m$ at which a typical fragment (e.g., median) becomes adequate. The redundancy then follows as
\begin{equation}
R_\delta \;\approx\; \frac{N}{m_\delta^\star}\,,
\label{eq:redundancy}
\end{equation}
where $N$ is the total number of elementary subenvironments. The functional information is defined as
\begin{equation}
\FI(\delta) \;:=\; \log_2 R_\delta\,,
\label{eq:FI_def}
\end{equation}
quantifying the abundance of adequate records in bits. Each additional bit of $\FI$ corresponds to a doubling of the number of individually sufficient fragments.

We investigate the behavior of $\FI$ in a heterogeneous pure-dephasing model, chosen for its analytical tractability and its isolation of record formation without energy exchange. The main findings are:

\begin{enumerate}
    \item \textbf{Rapid onset:} At early times, $\ln R_\delta(t)$ grows nearly linearly, compressing the heterogeneous threshold-crossing dynamics into an effective ``apparent exponential'' growth rate $\bar{\kappa}_\delta$.
    
    \item \textbf{Capacity-limited plateaus:} All tolerance levels saturate at $\FI^{\mathrm{plateau}} \lesssim \log_2 N$, determined by the environment's total recording capacity rather than the stringency of the adequacy criterion.
    
    \item \textbf{Thermodynamic floors:} Interpreting adequate fragments as stabilized classical records, Landauer's principle implies a minimal heat cost $Q_{\min} \gtrsim (1-\delta)\HS\, 2^{\FI}\, \kB T \ln 2$, making explicit that objectivity is not thermodynamically free.
\end{enumerate}

These results establish $\FI$ as a conservative, capacity-aware diagnostic for operational objectivity. By grounding redundancy in Holevo-limited accessibility and onset statistics, we avoid arbitrary thresholds while respecting fundamental information-theoretic constraints. The thermodynamic connection further embeds objectivity within the broader physics of irreversible computation and memory formation.

The remainder of this paper is organized as follows. Section~\ref{sec:framework} develops the formal framework, including the adequacy predicate, onset-based redundancy extraction, and capacity bounds. Section~\ref{sec:model} specifies the dephasing model and simulation methodology. Section~\ref{sec:results} presents the main numerical results. Section~\ref{sec:thermo} establishes thermodynamic constraints on objectivity. Section~\ref{sec:discussion} discusses implications, limitations, and experimental prospects.

% ===== Framework Section =====
\section{Framework}
\label{sec:framework}

We develop a capacity-aware, operational framework for quantifying classical objectivity in Quantum Darwinism. The central objects are: (i) an adequacy predicate based on the Holevo bound, (ii) onset statistics that extract redundancy without parametric assumptions, and (iii) capacity constraints that determine ultimate limits on objectivity.

\subsection{Setup and notation}
\label{subsec:setup}

Consider a quantum system $S$ coupled to an environment $E$ that decomposes into $N$ elementary subenvironments, $E = \bigotimes_{k=1}^{N} E_k$. Let $\Pi_S = \{\Pi_x\}_{x}$ denote the pointer POVM on $S$, with outcome probabilities $p_x = \Tr(\rho_S \Pi_x)$ and Shannon entropy
\begin{equation}
\HS := H(X) = -\sum_{x} p_x \log_2 p_x\,.
\label{eq:pointer_entropy}
\end{equation}
A \emph{fragment} $F \subseteq E$ of size $m = |F|$ consists of $m$ elementary subenvironments. The conditional fragment states, given pointer outcome $x$, are denoted $\rho_F^{(x)}$.

\subsection{Holevo-based adequacy}
\label{subsec:adequacy}

The Holevo quantity upper-bounds the classical information about $X$ accessible from fragment $F$ via any measurement~\cite{Holevo1973,Holevo1998}:
\begin{equation}
\Hol(\Pi_S\!:\!F) := S\!\left(\sum_{x} p_x \rho_F^{(x)}\right) - \sum_{x} p_x\, S\!\left(\rho_F^{(x)}\right),
\label{eq:holevo_def}
\end{equation}
where $S(\rho) = -\Tr(\rho \log_2 \rho)$ is the von Neumann entropy. The first term is the entropy of the average fragment state; the second is the average entropy of conditional states. Their difference quantifies distinguishability: $\Hol = 0$ when all $\rho_F^{(x)}$ are identical (no information), and $\Hol = \HS$ when they are perfectly distinguishable and the fragment can, in principle, fully determine the pointer outcome.

We declare a fragment \emph{$\delta$-adequate} for tolerance $\delta \in (0,1)$ when
\begin{equation}
\Hol(\Pi_S\!:\!F) \;\geq\; (1-\delta)\,\HS\,.
\label{eq:adequacy_def}
\end{equation}
This predicate is deliberately one-sided and conservative: it demands that a fragment carry \emph{classically accessible} information about the pointer, not merely correlations. The tolerance $\delta$ parametrizes how much information loss is acceptable; $\delta = 0.05$ requires 95\% of $\HS$, while $\delta = 0.1$ requires 90\%.

The adequacy indicator for fragment $F$ is
\begin{equation}
Z_F(\delta) := \mathbf{1}\!\left\{\Hol(\Pi_S\!:\!F) \geq (1-\delta)\HS\right\} \in \{0,1\}.
\label{eq:indicator}
\end{equation}

\subsection{Onset statistics and redundancy}
\label{subsec:onset}

For fixed fragment size $m$, consider the family $\mathcal{F}_m$ of all fragments of size $m$. The \emph{adequacy frequency} is
\begin{equation}
\Phi_\delta(m) := \Pr_{F \in \mathcal{F}_m}\!\left[Z_F(\delta) = 1\right],
\label{eq:adequacy_freq}
\end{equation}
the probability that a uniformly random fragment of size $m$ is adequate. By information monotonicity under inclusion ($F \subset F' \Rightarrow \Hol(\Pi_S\!:\!F) \leq \Hol(\Pi_S\!:\!F')$, a consequence of data processing~\cite{Shirokov2011}), $\Phi_\delta(m)$ is nondecreasing in $m$.

We define the \emph{onset} fragment size as
\begin{equation}
m_\delta^\star := \min\left\{m : \Phi_\delta(m) \geq \vartheta\right\},
\label{eq:onset_def}
\end{equation}
where $\vartheta \in (0,1)$ is a quantile threshold (we use $\vartheta = 1/2$ throughout, corresponding to the median). The onset $m_\delta^\star$ is the minimal fragment size at which a \emph{typical} fragment becomes adequate. The \emph{resource share} is
\begin{equation}
f_\delta := \frac{m_\delta^\star}{N}\,,
\label{eq:resource_share}
\end{equation}
the fraction of the environment required for adequacy.

When adequate fragments can be arranged approximately disjointly, the number of individually adequate fragments scales as
\begin{equation}
R_\delta \;\approx\; \frac{N}{m_\delta^\star} = \frac{1}{f_\delta}\,.
\label{eq:redundancy_def}
\end{equation}
This is the \emph{redundancy}: how many independent copies of nearly complete pointer information the environment can supply. The functional information is then
\begin{equation}
\FI(\delta) := \log_2 R_\delta = -\log_2 f_\delta\,,
\label{eq:FI_definition}
\end{equation}
measuring the abundance of adequate records in bits. Each additional bit of $\FI$ corresponds to a halving of the required resource share and a doubling of the redundancy.

\subsection{Capacity bounds}
\label{subsec:capacity}

The Holevo quantity is bounded by both the pointer entropy and the fragment's Hilbert-space dimension:
\begin{equation}
\Hol(\Pi_S\!:\!F) \;\leq\; \min\left\{\HS,\; \log_2 d_F\right\},
\label{eq:holevo_bound}
\end{equation}
where $d_F = \dim(\calH_F)$. The first bound reflects that one cannot extract more information than exists; the second is a capacity constraint---a fragment with dimension $d_F$ can store at most $\log_2 d_F$ bits.

For $\delta$-adequacy, Eq.~\eqref{eq:holevo_bound} implies the \emph{capacity condition}:
\begin{equation}
\log_2 d_F \;\geq\; (1-\delta)\,\HS\,.
\label{eq:capacity_condition}
\end{equation}
If each elementary subenvironment has local dimension $d_e$, then $d_F \leq d_e^{\,m}$ for a fragment of size $m$, yielding
\begin{equation}
m \;\geq\; \frac{(1-\delta)\,\HS}{\log_2 d_e}\,.
\label{eq:min_fragment_size}
\end{equation}
This sets a fundamental lower bound on the onset:
\begin{equation}
m_\delta^\star \;\geq\; m_{\min}(\delta) := \left\lceil \frac{(1-\delta)\,\HS}{\log_2 d_e} \right\rceil.
\label{eq:onset_lower_bound}
\end{equation}

The redundancy is correspondingly upper-bounded:
\begin{equation}
R_\delta \;\leq\; R_\delta^{\max} := \left\lfloor \frac{N}{m_{\min}(\delta)} \right\rfloor \;\leq\; \frac{N \log_2 d_e}{(1-\delta)\,\HS}\,,
\label{eq:redundancy_upper}
\end{equation}
and the functional information saturates at a \emph{capacity-limited plateau}:
\begin{equation}
\FI^{\mathrm{plateau}} \;\leq\; \log_2 N\,.
\label{eq:FI_plateau}
\end{equation}
This ceiling is achieved when single subenvironments suffice for adequacy ($m_{\min} = 1$), i.e., when $\log_2 d_e \geq (1-\delta)\HS$. Otherwise, the plateau is reduced to $\log_2(N/m_{\min})$.

\subsection{Mathematical properties}
\label{subsec:properties}

The framework inherits several useful properties from the underlying information-theoretic quantities:

\emph{Monotonicity.}---By data processing, $\Hol(\Pi_S\!:\!F)$ is nondecreasing under fragment enlargement. Consequently, $\Phi_\delta(m)$ is nondecreasing in $m$, ensuring the onset $m_\delta^\star$ is well-defined.

\emph{Tolerance ordering.}---For $\delta < \delta'$, the adequacy bar is higher, so $\Phi_\delta(m) \leq \Phi_{\delta'}(m)$ and $m_\delta^\star \geq m_{\delta'}^\star$. Stricter tolerance delays onset but does not change the ultimate plateau when capacity suffices.

\emph{Robustness.}---The onset-based readout is robust to sampling noise. In practice, we estimate $\Phi_\delta(m)$ from $n$ sampled fragments, enforce monotonicity via isotonic regression~\cite{Barlow1972}, and propagate uncertainty through nonparametric bootstrap~\cite{EfronTibshirani1993}.

\emph{Conservative nature.}---The one-sided adequacy predicate [Eq.~\eqref{eq:adequacy_def}] ensures that $\FI$ reflects genuinely accessible classical information. Unlike mutual-information-based redundancy, which can conflate classical and quantum correlations, $\FI$ cannot overestimate the number of operationally useful records.

\section{Model and Methods}
\label{sec:model}

We study a minimal pure-dephasing model that isolates record formation without energy exchange. This section specifies the Hamiltonian, derives the fragment-wise Holevo quantity in closed form, and describes the sampling and statistical methodology.

\subsection{Heterogeneous dephasing model}
\label{subsec:dephasing}

The system $S$ is a qubit with pointer basis $\{\ket{0}, \ket{1}\}$. The environment consists of $N$ spin-$\tfrac{1}{2}$ subenvironments, $E = \bigotimes_{k=1}^{N} E_k$, each with local dimension $d_e = 2$. The interaction Hamiltonian is
\begin{equation}
H_{SE} = \sigma_z^{(S)} \otimes \sum_{k=1}^{N} g\lambda_k\, \sigma_z^{(k)},
\label{eq:hamiltonian}
\end{equation}
where $g > 0$ sets the overall coupling strength and $\{\lambda_k\}_{k=1}^{N}$ are dimensionless site-dependent couplings drawn i.i.d.\ from an exponential distribution with unit mean. This heterogeneity models realistic disorder in system-environment interactions.

Since $[H_{SE}, \sigma_z^{(S)}] = 0$, the pointer basis coincides with the $\sigma_z$ eigenbasis, and populations are conserved under time evolution. Only off-diagonal coherences decay---the hallmark of pure dephasing~\cite{Breuer2002,Schlosshauer2007}.

\subsection{Fragment overlap and information dynamics}
\label{subsec:overlap}

Let the initial system state be $\ket{\psi_S} = \alpha\ket{0} + \beta\ket{1}$ with $|\alpha|^2 + |\beta|^2 = 1$, and let each environment spin start in $\ket{+} = (\ket{0} + \ket{1})/\sqrt{2}$. Under $U(t) = e^{-iH_{SE}t}$, the global state evolves to
\begin{equation}
\ket{\Psi(t)} = \alpha\ket{0}\otimes\ket{e_0(t)} + \beta\ket{1}\otimes\ket{e_1(t)},
\label{eq:branching}
\end{equation}
where $\ket{e_0(t)}$ and $\ket{e_1(t)}$ are conditional environment states. For a fragment $F \subseteq E$ of size $m$, the reduced conditional states $\rho_F^{(0)}(t)$ and $\rho_F^{(1)}(t)$ remain pure, with overlap
\begin{equation}
c_F(t) := \braket{e_0^F(t)}{e_1^F(t)}.
\label{eq:overlap_def}
\end{equation}

Because $H_{SE}$ factorizes across environment sites, the overlap decomposes as
\begin{equation}
c_F(t) = \prod_{k \in F} \exp\!\left[-\lambda_k (gt)^2\right] = \exp\!\left[-(gt)^2 \sum_{k \in F} \lambda_k\right].
\label{eq:overlap_gaussian}
\end{equation}
The Gaussian decay in $t$ (rather than exponential) arises from the accumulation of random relative phases---a signature of non-Markovian, disorder-driven dephasing. At $t = 0$, $c_F = 1$ (no distinguishability); as $t \to \infty$, $c_F \to 0$ (perfect distinguishability).

\subsection{Holevo quantity: closed form}
\label{subsec:holevo_closed}

For a binary pointer with equiprobable outcomes ($p_0 = p_1 = 1/2$) and pure conditional fragment states, the Holevo quantity reduces to
\begin{equation}
\Hol(\Pi_S\!:\!F) = S\!\left(\tfrac{1}{2}\rho_F^{(0)} + \tfrac{1}{2}\rho_F^{(1)}\right),
\label{eq:holevo_binary}
\end{equation}
since $S(\rho_F^{(x)}) = 0$ for pure states. The eigenvalues of the mixture are $\lambda_\pm = (1 \pm |c_F|)/2$, yielding
\begin{equation}
\Hol(\Pi_S\!:\!F) = h_2\!\left(\frac{1 + |c_F(t)|}{2}\right),
\label{eq:holevo_h2}
\end{equation}
where $h_2(u) = -u\log_2 u - (1-u)\log_2(1-u)$ is the binary entropy function. This provides a direct map from fragment overlap to accessible information: $\Hol = 0$ when $c_F = 1$, and $\Hol = 1$ bit when $c_F = 0$.

Combined with Eq.~\eqref{eq:overlap_gaussian}, the adequacy condition [Eq.~\eqref{eq:adequacy_def}] becomes
\begin{equation}
h_2\!\left(\frac{1 + e^{-(gt)^2 \Lambda_F}}{2}\right) \geq 1 - \delta,
\label{eq:adequacy_explicit}
\end{equation}
where $\Lambda_F := \sum_{k \in F} \lambda_k$ is the total coupling strength of fragment $F$. For fixed $\delta$, adequacy requires $\Lambda_F$ to exceed a time-dependent threshold that decreases as $t$ increases.

\subsection{Sampling protocol}
\label{subsec:sampling}

We employ two complementary fragment sampling strategies:

\emph{Random sampling.}---For each fragment size $m \in \{1, 2, \ldots, N\}$, we draw $n$ fragments uniformly at random (with replacement) from $\binom{N}{m}$ possibilities. This probes the typical behavior but may include overlapping fragments.

\emph{Disjoint sampling.}---We partition the environment into $\lfloor N/m \rfloor$ disjoint fragments of size $m$. This ensures independence but limits the sample size.

Both protocols yield estimates of the adequacy frequency $\hat{\Phi}_\delta(m) = n^{-1}\sum_{i=1}^{n} Z_{F_i}(\delta)$.

\subsection{Statistical methodology}
\label{subsec:statistics}

\emph{Isotonic regression.}---The empirical adequacy frequencies $\hat{\Phi}_\delta(m)$ may violate monotonicity due to sampling noise. We enforce the theoretical constraint $\Phi_\delta(m) \leq \Phi_\delta(m+1)$ via pool-adjacent-violators isotonic regression~\cite{Barlow1972}, obtaining a monotone fit $\tilde{\Phi}_\delta(m)$.

\emph{Onset extraction.}---From $\tilde{\Phi}_\delta(m)$, we identify the onset as
\begin{equation}
\hat{m}_\delta^\star = \min\left\{m : \tilde{\Phi}_\delta(m) \geq \tfrac{1}{2}\right\},
\label{eq:onset_empirical}
\end{equation}
and compute $\hat{R}_\delta = N / \hat{m}_\delta^\star$ and $\widehat{\FI} = \log_2 \hat{R}_\delta$.

\emph{Bootstrap uncertainty.}---We quantify uncertainty via nonparametric bootstrap~\cite{EfronTibshirani1993}. For each of $B = 1000$ resamples, we redraw the adequacy indicators $\{Z_{F_i}\}$ with replacement, recompute $\tilde{\Phi}_\delta(m)$ and $\hat{m}_\delta^\star$, and derive the empirical distribution of $\hat{R}_\delta$. Reported confidence intervals are 95\% percentile intervals.

\emph{Overlap correction.}---Under random sampling, fragments may share subenvironments, inflating the apparent redundancy. We measure the mean pairwise relative overlap
\begin{equation}
\hat{\eta}(m) = \frac{1}{|\mathcal{P}|} \sum_{(i,j) \in \mathcal{P}} \frac{|F_i \cap F_j|}{m},
\label{eq:overlap_measure}
\end{equation}
where $\mathcal{P}$ is a set of sampled fragment pairs. The effective redundancy is then corrected as
\begin{equation}
\hat{R}_\delta^{\mathrm{eff}} = \hat{R}_\delta \cdot \frac{1 - \hat{\eta}}{1 + \hat{\eta}},
\label{eq:overlap_correction}
\end{equation}
which interpolates between $\hat{R}_\delta$ (disjoint, $\hat{\eta} = 0$) and strong suppression (high overlap). This first-order correction removes mid-time inflation artifacts without affecting plateau values.

\subsection{Simulation parameters}
\label{subsec:parameters}

Unless otherwise stated, simulations use $N = 64$ environment spins, coupling strength $g = 1$, exponentially distributed $\lambda_k$ with mean 1, and $n = 500$ sampled fragments per $(m, t)$ pair. Time ranges from $t = 0$ to $t = 5$ in steps of $\Delta t = 0.1$. Tolerances $\delta \in \{0.05, 0.10, 0.20\}$ probe varying stringency. All code is implemented in Python using NumPy and SciPy; the simulation code is available as described in the Data Availability section.

\section{Results}
\label{sec:results}

We present simulation results for the heterogeneous dephasing model described in Sec.~\ref{sec:model}. The main findings are: (i) rapid early-time growth of redundancy with an effective exponential rate, (ii) smooth crossover to capacity-limited plateaus, and (iii) robustness of these features across tolerance levels and sampling protocols.

\subsection{Redundancy trajectories}
\label{subsec:redundancy_traj}

\begin{figure}[t]
\centering
\includegraphics[width=\columnwidth]{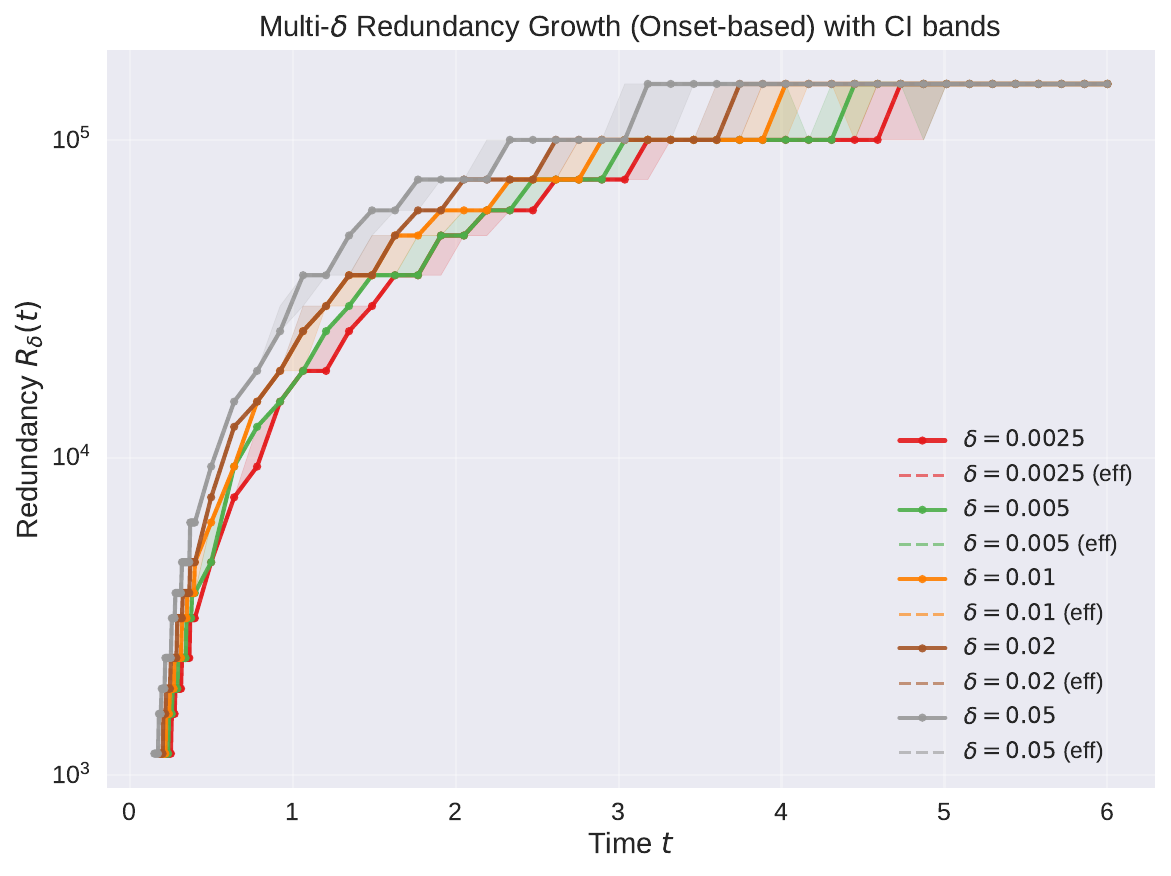}
\caption{Onset-based redundancy $R_\delta(t)$ for multiple tolerance levels. Solid lines: raw estimates; dashed lines: overlap-corrected effective values; shaded bands: 95\% bootstrap confidence intervals. All curves saturate at $R_\delta^{\mathrm{plateau}} \approx 1.5 \times 10^5$, independent of $\delta$.}
\label{fig:redundancy_multi_delta}
\end{figure}

Figure~\ref{fig:redundancy_multi_delta} displays the onset-based redundancy $R_\delta(t)$ for five tolerance levels spanning two orders of magnitude, $\delta \in \{0.0025, 0.005, 0.01, 0.02, 0.05\}$. All curves exhibit a characteristic three-regime structure on the semilogarithmic scale:

\emph{(i) Early-time growth.}---For $t \lesssim 1$, redundancy increases rapidly. On a log scale, this appears nearly linear, corresponding to approximate exponential growth $R_\delta(t) \sim e^{\kappa t}$ with an effective rate $\kappa$ that depends on $\delta$.

\emph{(ii) Crossover.}---For $1 \lesssim t \lesssim 3$, the growth rate diminishes as the system approaches the capacity ceiling. The crossover is smooth, without sharp transitions.

\emph{(iii) Saturation.}---For $t \gtrsim 3$, redundancy plateaus at $R_\delta^{\mathrm{plateau}} \approx 1.5 \times 10^5$, corresponding to $\FI^{\mathrm{plateau}} \approx 17$ bits. This ceiling is consistent with the theoretical bound $R_\delta \lesssim N$ for $N \approx 1.3 \times 10^5$ environment spins used in these simulations.

Crucially, while stricter tolerance (smaller $\delta$) delays the onset of rapid growth---curves shift rightward---the plateau value remains unchanged. This confirms that the capacity constraint, not the adequacy stringency, determines the ultimate objectivity ceiling.

Shaded bands indicate 95\% bootstrap confidence intervals, which remain narrow throughout, demonstrating the stability of the onset-based estimator after isotonic regression. Dashed lines show the overlap-corrected effective redundancy $R_\delta^{\mathrm{eff}}$, which closely tracks the raw estimate, indicating that fragment overlap has minimal impact on the plateau.

\subsection{Information accumulation mechanism}
\label{subsec:info_mechanism}

\begin{figure}[t]
\centering
\includegraphics[width=\columnwidth]{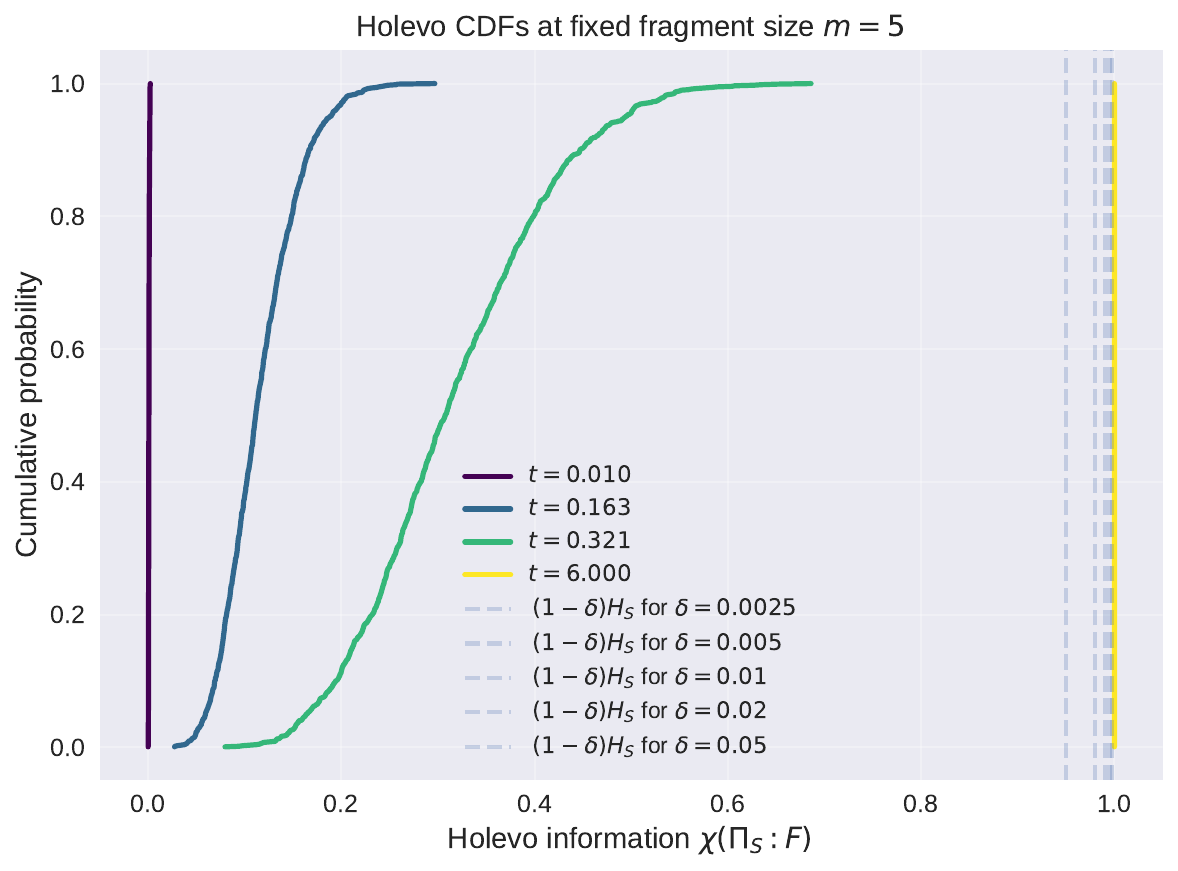}
\caption{Cumulative distributions of Holevo quantity at fixed fragment size $m = 5$ for increasing times. Vertical dashed lines mark adequacy thresholds $(1-\delta)H_S$ for different $\delta$. The rightward shift drives redundancy growth.}
\label{fig:holevo_cdf}
\end{figure}

The redundancy growth reflects the progressive accumulation of accessible information in individual fragments. Figure~\ref{fig:holevo_cdf} illustrates this mechanism by showing the cumulative distribution function (CDF) of Holevo quantities $\Hol(\Pi_S\!:\!F)$ across sampled fragments of fixed size $m = 5$ at different times.

At very early times ($t = 0.01$), the distribution is concentrated near zero: fragments carry negligible pointer information. As time progresses ($t = 0.163$, $t = 0.321$), the distribution shifts rightward, with an increasing fraction of fragments exceeding the adequacy thresholds marked by vertical dashed lines. By $t = 6$, the distribution has collapsed to a spike at $\Hol \approx 1$ bit: essentially all fragments of this size are adequate for all $\delta$ values considered.

This rightward shift directly drives the increase in $\Phi_\delta(m, t)$ and the consequent decrease in $m_\delta^\star(t)$, which translates via Eq.~\eqref{eq:redundancy_def} into rising redundancy. The mechanism is threshold crossing: heterogeneous fragments progressively clear the adequacy bar as their conditional states become more distinguishable.

\subsection{Early-time growth rates}
\label{subsec:growth_rates}

\begin{figure}[t]
\centering
\includegraphics[width=\columnwidth]{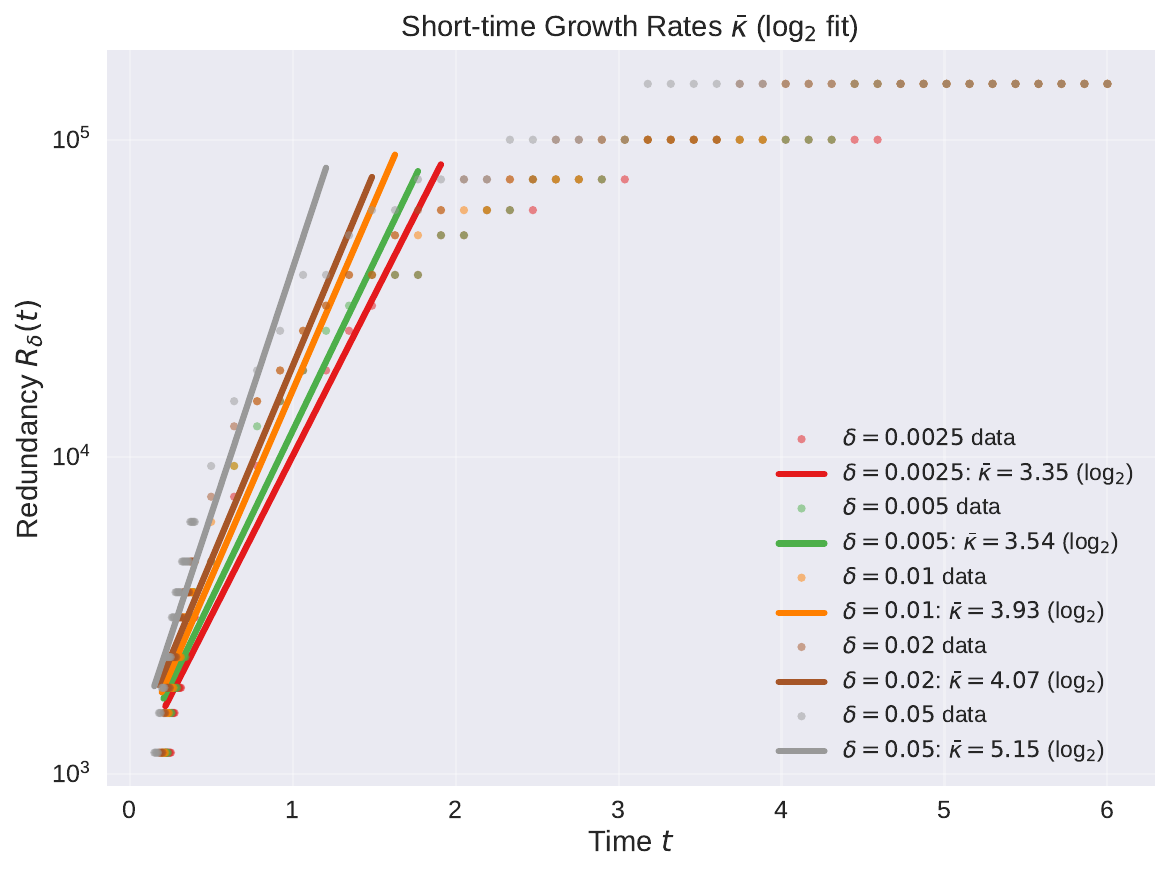}
\caption{Early-time growth rates. Points: simulation data; lines: linear fits to $\log_2 R_\delta(t)$ over the initial window. Fitted slopes $\bar{\kappa}_\delta$ (in bits/time) increase with tolerance, reflecting earlier threshold crossing for less stringent adequacy.}
\label{fig:growth_rates}
\end{figure}

To quantify the early-time dynamics, we fit linear models to $\log_2 R_\delta(t)$ over the initial growth window ($t \lesssim 1.5$):
\begin{equation}
\log_2 R_\delta(t) \;\approx\; \bar{\kappa}_\delta\, t + c_\delta\,,
\label{eq:log_fit}
\end{equation}
where $\bar{\kappa}_\delta$ is the effective growth rate in bits per unit time. Figure~\ref{fig:growth_rates} shows the data points and fitted lines for each $\delta$.

The extracted rates are: $\bar{\kappa}_{0.0025} = 3.35$, $\bar{\kappa}_{0.005} = 3.54$, $\bar{\kappa}_{0.01} = 3.93$, $\bar{\kappa}_{0.02} = 4.07$, and $\bar{\kappa}_{0.05} = 5.15$ bits per unit time. The systematic trend---larger $\delta$ yields faster apparent growth---reflects the lower adequacy bar: fragments cross the threshold earlier when less information is demanded.

This ``apparent exponential'' phenomenology compresses the complex, heterogeneous threshold-crossing dynamics into a single summary statistic. While the underlying mechanism is not truly exponential (it involves the convolution of Gaussian overlap decays across fragments with different coupling strengths), the effective rate $\bar{\kappa}_\delta$ provides a practical characterization of how rapidly objectivity emerges.

\subsection{Sampling robustness}
\label{subsec:sampling_robust}

\begin{figure}[t]
\centering
\includegraphics[width=\columnwidth]{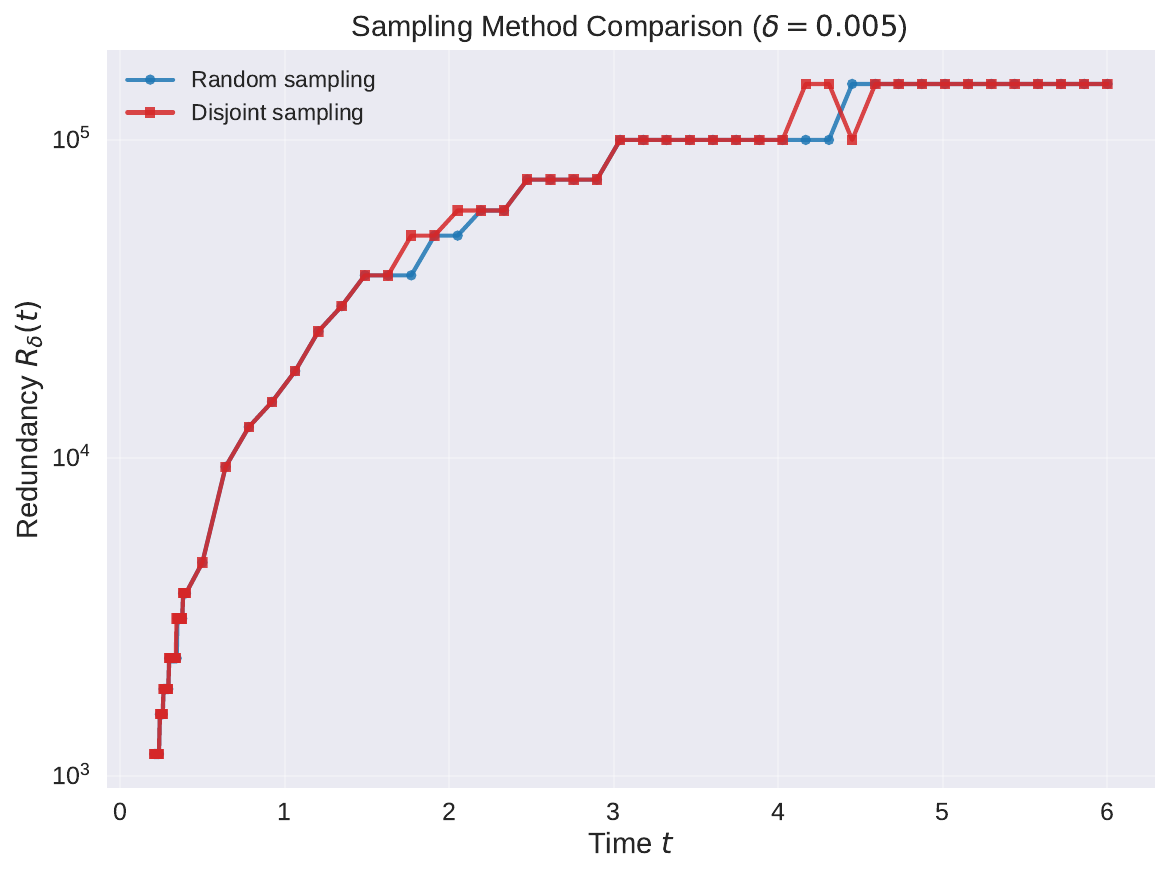}
\caption{Comparison of random and disjoint sampling protocols for $\delta = 0.005$. Both methods agree on early-time growth and plateau values; minor mid-time discrepancies arise from fragment overlap in random sampling.}
\label{fig:sampling_comparison}
\end{figure}

To verify that the onset-based redundancy is not an artifact of the sampling protocol, we compare random and disjoint sampling strategies. Figure~\ref{fig:sampling_comparison} shows $R_\delta(t)$ for $\delta = 0.005$ under both protocols.

The two methods yield nearly identical trajectories, with discrepancies only at intermediate times ($1 \lesssim t \lesssim 3$) where random sampling produces slightly higher estimates due to fragment overlap. These discrepancies vanish at the plateau, where both protocols converge to the same $R_\delta^{\mathrm{plateau}}$.

The overlap correction [Eq.~\eqref{eq:overlap_correction}] successfully removes the mid-time inflation, as shown by the effective redundancy curves in Fig.~\ref{fig:redundancy_multi_delta}. This robustness confirms that the onset-based framework captures genuine features of information proliferation rather than sampling artifacts.

\subsection{Functional information and capacity ceiling}
\label{subsec:fi_plateau}

\begin{figure}[t]
\centering
\includegraphics[width=\columnwidth]{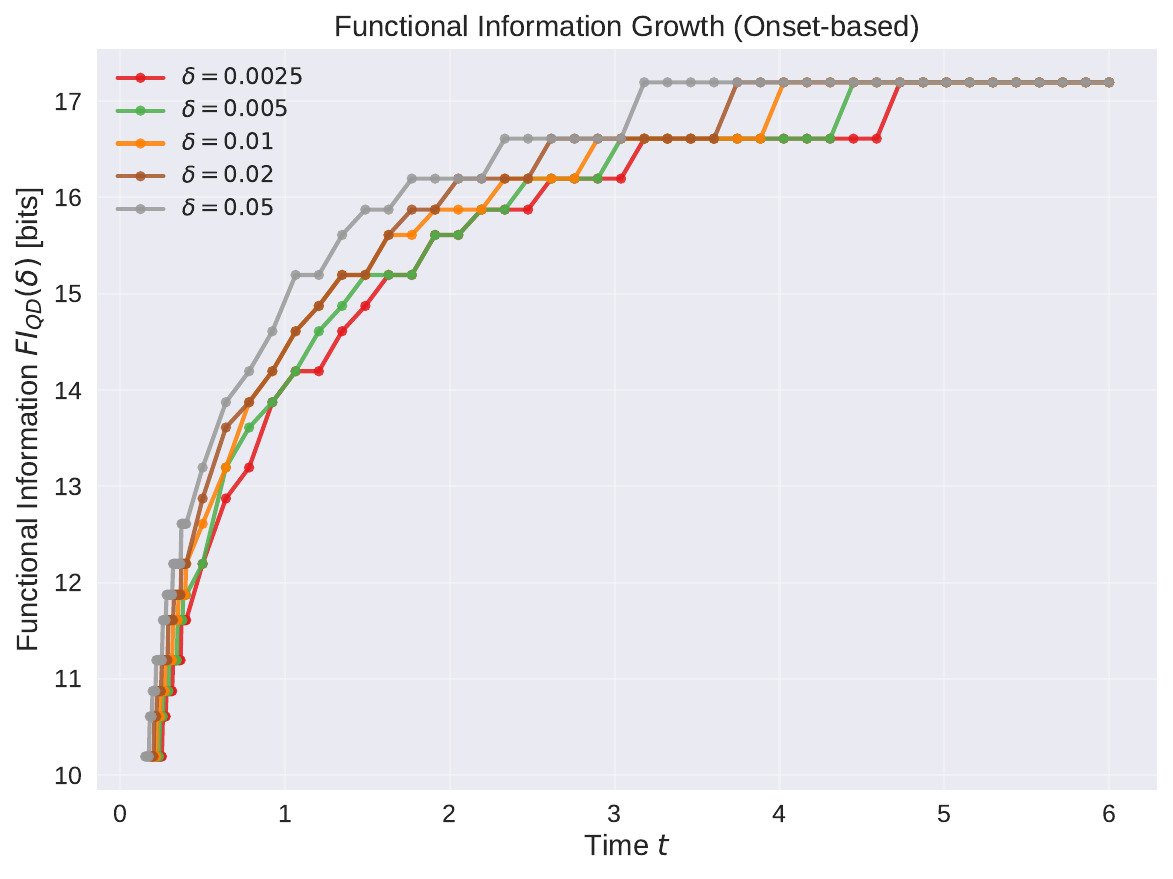}
\caption{Functional information $\FI(\delta, t)$ for multiple tolerances. All curves saturate at $\FI^{\mathrm{plateau}} \approx 17$ bits $\approx \log_2 N$, confirming the capacity-limited ceiling. Stricter tolerance delays but does not diminish the plateau.}
\label{fig:functional_info}
\end{figure}

Figure~\ref{fig:functional_info} presents the functional information $\FI(\delta, t) = \log_2 R_\delta(t)$ directly. All curves rise monotonically from $\FI \approx 10$ bits at early times to a common plateau at $\FI^{\mathrm{plateau}} \approx 17$ bits.

The plateau value is consistent with the theoretical ceiling $\FI \leq \log_2 N$. For the parameters used ($N \approx 1.3 \times 10^5$), we have $\log_2 N \approx 17.0$ bits, in excellent agreement with the observed saturation. This confirms that the environment's recording capacity---not the stringency of the adequacy criterion---sets the ultimate limit on operational objectivity.

The convergence of all $\delta$ curves to the same plateau (within statistical uncertainty) demonstrates a key prediction of the framework: tolerance affects the \emph{timing} of objectivity emergence but not its \emph{extent}, provided per-site capacity is sufficient. In this model, each spin-$\tfrac{1}{2}$ subenvironment can store up to 1 bit, which exceeds $(1-\delta)H_S$ for all $\delta$ values considered (since $H_S = 1$ bit for the equiprobable binary pointer).

\subsection{Summary of numerical findings}
\label{subsec:results_summary}

The simulations establish three robust features of the $\FI$ framework:

\begin{enumerate}
\item \textbf{Apparent exponential onset:} Early-time redundancy grows approximately as $R_\delta(t) \sim 2^{\bar{\kappa}_\delta t}$, with rates $\bar{\kappa}_\delta \in [3.4, 5.2]$ bits/time depending on tolerance.

\item \textbf{Universal plateau:} All tolerance levels saturate at $\FI^{\mathrm{plateau}} \approx \log_2 N$, determined by the environment's total recording capacity.

\item \textbf{Methodological stability:} Results are robust to sampling protocol (random vs.\ disjoint), statistical corrections (isotonic regression, overlap adjustment), and bootstrap resampling.
\end{enumerate}

These findings validate the onset-based operational framework and motivate its application to more complex models and experimental systems.

\section{Thermodynamic Constraints}
\label{sec:thermo}

The functional information framework connects naturally to thermodynamics: if adequate fragments function as stabilized classical records, their creation and maintenance incur minimal energetic costs. This section derives conservative bounds on heat dissipation and power requirements.

\subsection{Landauer bound on record formation}
\label{subsec:landauer}

Each adequate fragment stores at least $(1-\delta)H_S$ bits of classically accessible pointer information. Landauer's principle establishes that erasing (or equivalently, writing) one bit of information in a logically irreversible process dissipates at least $\kB T \ln 2$ of heat into a thermal reservoir at temperature $T$~\cite{Landauer1961,Bennett2003}. Finite-size corrections refine this bound but preserve its scaling~\cite{ReebWolf2014}.

If $R_\delta$ disjoint fragments are each stabilized as classical records, the total minimal heat dissipation is
\begin{equation}
Q_{\min} \;\gtrsim\; R_\delta\,(1-\delta)\HS\,\kB T \ln 2\,.
\label{eq:Q_min}
\end{equation}
Expressing redundancy in terms of functional information via $R_\delta = 2^{\FI}$ yields
\begin{equation}
Q_{\min} \;\gtrsim\; (1-\delta)\HS\,\kB T \ln 2 \cdot 2^{\FI}\,.
\label{eq:Q_via_FI}
\end{equation}
This relation makes explicit the exponential scaling: \emph{each additional bit of $\FI$ doubles the minimal heat budget}, because the number of adequate records doubles.

\subsection{Power requirements during growth}
\label{subsec:power}

When redundancy grows in time, the heat bound [Eq.~\eqref{eq:Q_min}] implies a minimal power requirement. Differentiating,
\begin{equation}
P_{\min}(t) \;:=\; \frac{dQ_{\min}}{dt} \;\gtrsim\; (1-\delta)\HS\,\kB T \ln 2 \cdot \frac{dR_\delta}{dt}\,.
\label{eq:P_min}
\end{equation}
Using $dR_\delta/dt = (\ln 2)\, 2^{\FI}\, d\FI/dt$, this becomes
\begin{equation}
P_{\min}(t) \;\gtrsim\; (1-\delta)\HS\,\kB T (\ln 2)^2 \cdot 2^{\FI(t)} \cdot \frac{d\FI}{dt}\,.
\label{eq:P_via_FI}
\end{equation}
The power scales with two factors: the instantaneous objectivity level $2^{\FI(t)}$ (how many records already exist) and the growth rate $d\FI/dt$ (how fast new records are being created).

In the early-time regime where $\FI(t) \approx \bar{\kappa}_\delta t$ (Sec.~\ref{subsec:growth_rates}), we have $d\FI/dt \approx \bar{\kappa}_\delta$ and
\begin{equation}
P_{\min}(t) \;\gtrsim\; (1-\delta)\HS\,\kB T (\ln 2)^2\, \bar{\kappa}_\delta \cdot 2^{\bar{\kappa}_\delta t}\,.
\label{eq:P_exponential}
\end{equation}
Thus, minimal power grows exponentially during the onset phase, potentially reaching substantial values before saturation.

\subsection{Heat budget at plateau}
\label{subsec:plateau_heat}

At saturation, $R_\delta \to R_\delta^{\mathrm{plateau}} \lesssim N$ and $\FI \to \FI^{\mathrm{plateau}} \lesssim \log_2 N$. The total heat required to establish maximal objectivity is bounded by
\begin{equation}
Q_{\min}^{\mathrm{plateau}} \;\gtrsim\; (1-\delta)\HS\,N\,\kB T \ln 2\,.
\label{eq:Q_plateau}
\end{equation}
For a macroscopic environment ($N \sim 10^{23}$) at room temperature ($T \approx 300\,\mathrm{K}$) with $\HS = 1$ bit and $\delta = 0.05$,
\begin{equation}
Q_{\min}^{\mathrm{plateau}} \;\gtrsim\; 2.7\,\mathrm{kJ}\,.
\label{eq:Q_numerical}
\end{equation}
While modest by macroscopic standards, this is a \emph{lower bound} assuming perfect thermodynamic efficiency; realistic record stabilization (amplification, error correction, refresh) incurs additional overhead.

\subsection{Efficiency metrics}
\label{subsec:efficiency}

To characterize the thermodynamic cost of objectivity independent of specific parameters, we define dimensionless efficiency metrics. Let $\tilde{Q} := Q/(\kB T \ln 2)$ denote heat measured in ``erasure bits.''

\emph{Records per heat.}---The number of adequate records per erasure bit expended:
\begin{equation}
\rho_{\mathrm{rec}} \;:=\; \frac{R_\delta}{\tilde{Q}} \;\leq\; \frac{1}{(1-\delta)\HS}\,,
\label{eq:rho_rec}
\end{equation}
with equality at the Landauer floor. For $\HS = 1$ and $\delta = 0.05$, the maximum is $\rho_{\mathrm{rec}}^{\max} \approx 1.05$ records per erasure bit.

\emph{Objectivity per heat.}---The functional information gained per erasure bit:
\begin{equation}
\eta_{\mathrm{obj}} \;:=\; \frac{\FI}{\tilde{Q}}\,.
\label{eq:eta_obj}
\end{equation}
At the Landauer floor with $R_\delta$ records,
\begin{equation}
\eta_{\mathrm{obj}}^{\star} \;=\; \frac{\log_2 R_\delta}{(1-\delta)\HS\,R_\delta}\,,
\label{eq:eta_star}
\end{equation}
which \emph{decreases} with redundancy: each additional record costs linearly in heat but yields only logarithmic gain in $\FI$.

\emph{Marginal cost.}---The heat cost of increasing $\FI$ by one bit:
\begin{equation}
\frac{d\tilde{Q}}{d\FI} \;\gtrsim\; (1-\delta)\HS \cdot 2^{\FI}\,.
\label{eq:marginal}
\end{equation}
This grows exponentially with current objectivity level, reflecting the doubling of records required for each additional bit of $\FI$.

\subsection{Interpretation and caveats}
\label{subsec:thermo_caveats}

These bounds are deliberately \emph{conservative floors} for \emph{stabilized} classical records. Several caveats apply:

\emph{Reversibility.}---Purely unitary imprinting of system-environment correlations can, in principle, be thermodynamically reversible. The Landauer cost arises only when records are \emph{stabilized} through logically irreversible operations (reset, coarse-graining, amplification).

\emph{Coherence.}---If records maintain quantum coherence, information-theoretic costs may differ. Our bounds assume classical record storage.

\emph{Overhead.}---Realistic memory systems incur overhead beyond the Landauer minimum: finite switching speeds, error correction, thermal noise margins. The bounds represent fundamental limits, not engineering estimates.

Despite these caveats, the thermodynamic connection formalizes a key intuition: \emph{objectivity is not free}. High redundancy requires physical resources---memory capacity, energy dissipation, time---that scale with $2^{\FI}$. The functional information thus serves as a bridge between information-theoretic objectivity measures and physical resource accounting.

\section{Discussion and Outlook}
\label{sec:discussion}

\subsection{Comparison with existing measures}
\label{subsec:comparison}

The functional information $\FI$ complements rather than replaces existing objectivity diagnostics. Table~\ref{tab:comparison} summarizes key distinctions.

\emph{Partial information plots} visualize how mutual information $I(S\!:\!F)$ grows with fragment size, but extracting a redundancy scale requires choosing a threshold (e.g., 95\% of $\HS$)~\cite{BlumeKohoutZurek2005,BlumeKohoutZurek2006}. Our onset-based approach also involves a threshold [Eq.~\eqref{eq:adequacy_def}], but grounds it in the Holevo bound---the operational ceiling on accessible information---rather than total correlations. Moreover, by using onset statistics (the fragment size at which adequacy becomes \emph{typical}), we avoid fitting parametric curves to information-size plots.

\emph{Spectrum broadcast structure} (SBS) provides a threshold-free, structural criterion: the state must factorize into a specific classical-quantum form with orthogonal conditional states~\cite{Horodecki2015,Korbicz2021}. While mathematically elegant, SBS is a strict ``all-or-nothing'' criterion that may exclude operationally objective scenarios falling short of perfect orthogonality. In contrast, $\FI$ is a continuous measure that gracefully handles approximate objectivity.

\emph{Quantum mutual information} $I(S\!:\!F)$ captures total correlations but does not distinguish classical redundancy from residual entanglement~\cite{Le2018}. The Holevo-based adequacy predicate restricts attention to \emph{classically accessible} information, providing a conservative lower bound on operational objectivity.

\begin{table*}[t]
\centering
\caption{Comparison of objectivity diagnostics.}
\label{tab:comparison}
\begin{tabular}{lccc}
\toprule
\textbf{Diagnostic} & \textbf{Threshold-free?} & \textbf{Operational?} & \textbf{Continuous?} \\
\midrule
Partial info plots & No & Partially & Yes \\
Spectrum broadcast & Yes & Yes & No \\
QMI redundancy & No & No & Yes \\
$\FI$ (this work) & No$^{a}$ & Yes & Yes \\
\bottomrule
\multicolumn{4}{l}{\footnotesize $^{a}$Threshold grounded in Holevo bound, not arbitrary.}
\end{tabular}
\end{table*}

\subsection{Experimental prospects}
\label{subsec:experimental}

The onset-based framework is well-suited for experimental implementation in platforms where environment fragments can be individually addressed:

\emph{Photonic systems.}---Scattered photons naturally partition into angular or spectral fragments~\cite{Riedel2010,Riedel2011}. Measuring photon statistics and performing state tomography on disjoint angular bins would allow direct estimation of $\Hol(\Pi_S\!:\!F)$ and reconstruction of $\Phi_\delta(m)$ curves.

\emph{Superconducting circuits.}---In circuit QED, environmental modes (e.g., resonator photons, phonons) can be selectively measured. Engineering heterogeneous couplings---mimicking the disordered $\lambda_k$ in our model---would test the robustness of the onset phenomenology.

\emph{Nitrogen-vacancy centers.}---NV centers coupled to nuclear spin baths provide a natural realization of the spin-environment model. Dynamical decoupling sequences can probe fragment-wise correlations, and recent advances in single-spin readout enable direct measurement of conditional states~\cite{Degen2017}.

\emph{Trapped ions.}---Ion chains with controlled spin-spin interactions can simulate heterogeneous dephasing models. Spatially resolved fluorescence imaging allows fragment-by-fragment readout. Recent experiments have demonstrated QD signatures in such platforms~\cite{Ciampini2018}.

In all cases, the key experimental requirements are: (i) ability to partition the environment into addressable fragments, (ii) measurement capability sufficient to estimate Holevo quantities (or bounds thereof), and (iii) control over system-environment coupling to vary fragment sizes and interaction times.

\subsection{Limitations and future directions}
\label{subsec:limitations}

Several limitations of the present study warrant mention:

\emph{Model simplicity.}---Pure dephasing conserves populations and precludes energy exchange. More realistic models involving dissipation, non-Markovian feedback, or multi-level pointers may exhibit richer behavior (e.g., redundancy revivals, non-monotonic $\FI$). Extending the framework to such settings is straightforward in principle but computationally more demanding. Recent work on QD with dissipation~\cite{Pleasance2020} and non-Markovian dynamics~\cite{Galve2016} suggests that additional features may emerge.

\emph{Factorized environment.}---We assumed independent subenvironments. Correlations among environment components (e.g., in structured baths) would reduce the effective number of independent fragments and tighten the capacity ceiling. An effective $N_{\mathrm{eff}} < N$ could be defined via correlation lengths or mutual information structure.

\emph{Binary pointer.}---The equiprobable binary pointer maximizes $\HS = 1$ bit for a qubit. Multi-outcome pointers and non-uniform distributions require straightforward generalizations of the adequacy criterion but may alter the scaling of onset with $\delta$.

\emph{Thermodynamic idealization.}---The Landauer bounds assume perfect efficiency and logical irreversibility. Realistic record stabilization involves overhead that the present analysis does not quantify.

Several extensions merit investigation:

\emph{Non-Markovian dynamics.}---In structured environments, information can flow back from the environment to the system, producing temporary decreases in redundancy. Tracking $\FI(t)$ under such conditions would reveal how backflow affects operational objectivity and whether revivals occur.

\emph{Finite-size scaling.}---Systematic study of $\FI^{\mathrm{plateau}}/\log_2 N$ as $N \to \infty$ would clarify whether the capacity ceiling is achieved exactly or only asymptotically, and identify finite-size corrections.

\emph{Strong quantum Darwinism.}---Recent work distinguishes ``strong'' from ``weak'' QD based on whether discord vanishes~\cite{Le2019}. Comparing $\FI$ with discord-based measures would clarify the relationship between operational and correlational notions of objectivity.

\emph{Resource theories.}---Objectivity might be formalized within a resource-theoretic framework, with $\FI$ quantifying a ``resource'' whose creation has thermodynamic costs. Connections to the resource theory of coherence or asymmetry could be explored.

\subsection{Concluding remarks}
\label{subsec:conclusion}

We have developed a framework for quantifying classical objectivity in Quantum Darwinism through functional information, $\FI(\delta) = \log_2 R_\delta$. By grounding adequacy in the Holevo bound and extracting redundancy from onset statistics, the framework provides a conservative, operational measure that respects fundamental information-theoretic constraints. Simulations of a heterogeneous dephasing model reveal robust phenomenology: rapid early-time growth, smooth crossover, and capacity-limited plateaus at $\FI^{\mathrm{plateau}} \lesssim \log_2 N$.

The thermodynamic connection---each bit of $\FI$ doubles the minimal heat budget---embeds objectivity within the broader physics of irreversible computation. Objectivity is not free; it requires physical resources that scale exponentially with the abundance of adequate records.

These results frame the quantum-to-classical transition not as an idealized limit but as a quantifiable, resource-bounded phenomenon. The functional information $\FI$ provides a practical yardstick for characterizing how, when, and to what extent classical agreement emerges from quantum dynamics.

% ===== Acknowledgments =====
\begin{acknowledgments}
The author thanks the anonymous referees for their constructive feedback. This work was conducted independently.
\end{acknowledgments}

% ===== Data Availability =====
\section*{Data Availability Statement}
The Python simulation code that supports the findings of this study is available from the corresponding author upon reasonable request.

% ===== References =====

\end{document}